\documentclass[12pt]{article}
\usepackage[dvips]{graphicx}
\usepackage{amssymb,amsmath,epsf}
\usepackage{epsfig}
\usepackage{wrapfig}
\def\pp{{\mbox{\boldmath$p$}}}
\def\p'{{\mbox{\boldmath$p'$}}}

\def\k{{\mbox{\boldmath$k$}}}

\def\q{{\mbox{\boldmath$q$}}}
\def\l{{\mbox{\boldmath$l$}}}

\def\zz{{\mbox{\boldmath$0$}}}
\def\re{{\mbox{\rm Re}}}

\begin{document}

\begin{center}
{\large{\bf ON THE ELECTRODISINTEGRATION OF THE DEUTERON IN THE
BETHE-SALPETER FORMALISM}}

\vspace*{5mm}

\underline{S.G. Bondarenko$^{a,}$}\footnote{E--mail: {\tt
bondarenko@jinr.ru}}, V.V. Burov$^{a}$, A.A. Goy$^{b}$, and E.P.
Rogochaya$^{a}$

\vspace*{3mm}

{\small
{\it a) JINR, 141980, Dubna, Moscow region, Russia} \\
{\it b) FENU, 690950, Vladivostok, Russia} }
\end{center}

\vspace*{5mm}

{\small{ \centerline{\bf Abstract} The $(ed\rightarrow enp)$
process in the frame of the Bethe-Salpeter approach with a
separable kernel of the Nucleon-Nucleon (NN) interaction was
considered. This conception keeps the covariance of description of
the process. Special attention was devoted to a contribution of
the $D$-states of the deuteron in the cross section of the
electrodisintegration. It was shown that the spectator particle
(neutron) plays an important role. The factorization of a cross
section of this reaction in the impulse approximation was checked
by analytical and numerical calculations.}}

\vspace*{3mm}

{\large\bf 1.~~Introduction}

\vspace*{3mm}

Study of static and dynamic electromagnetic properties of light
nuclei and, especially, the deuteron enables more deeply to
understand a nature of strong interactions and, in particular,
nucleon-nucleon interactions. At high energies considerations of a
nucleus as a nucleon system are not well justified. For this
reason the problems to study non-nucleonic degrees of freedom
(mesons, $ \Delta $-isobars, quark admixtures etc.) in an
intermediate energies region are widely discussed. However, in
spite of it significant progress was achieved on this way
relativistic effects (which {\em a priori} are very important at
large transfer momenta) are needed to include in the
consideration.

Other actively discussed problem is  the extraction from
experiments with light nuclei of the information about a structure
of bounded nucleons.
It requires to take into account relativistic kinematics of the
reaction and dynamics of interaction.
So the construction of a covariant approach and detailed analysis
of relativistic effects in electromagnetic reactions with light
nuclei are very important and interesting.

The electrodisintegration of the deuteron at the threshold has
been of interest of an investigation for a long time
\cite{arenh3}-\cite{gakh}. The reason is that the
electrodisintegration is an essential instrument for study a
structure of a two-nucleon system. First of all it is an
electromagnetic structure. The deuteron has been used as a neutron
target to get the information about neutron electromagnetic form
factors. During last 20 years it has also been used to receive
constraints on available realistic NN potentials. Analyzing of the
electrodisintegration process we can clarify the role of
non-nucleonic degrees of freedom. The non-nucleonic effects are
often important in few-body systems. And the deuteron is one of
convenient candidates because complete calculations can in
principle be performed.

First experiments were carried out at low transfer momenta but
more modern experiments \cite{bernheim}-\cite{boden} performed at
high momenta have opened many new questions in a region where
relativistic effects are important.  The experimental results on
the differential cross section derived from $(ed \rightarrow enp)$
reaction are available up to a momentum transfer of about 1 GeV.
This situation is very good for investigation of the deuteron
structure  at short distances with the allowance for some exotic
effects which have not been earlier important. First of all these
are the quark degrees of freedom (see \cite{kiss},
\cite{our-quarks}, for instance), but formerly it is necessary to
take into account relativistic effects.

Bethe-Salpeter (BS) approach~\cite{BS} can give a possibility to
consider relativistic effects by consistent way \cite{burov}. In
the paper the deuteron electrodisintegration within the covariant
BS approach with the separable Graz II interaction kernel is
presented. The exclusive differential cross section is calculated
in the plane wave relativistic impulse approximation.

The paper is organized as follows. In section 2 the relativistic
kinematics of the reaction and formulae for the cross
section are considered. The BS amplitude is presented in section 3.
The hadron current
in the BS formalism is defined in section 4. Factorization of the
cross section is discussed in section 5. Then the results of our
numerical calculations are presented in section 6. Finally the
discussion of the results is performed and further plans are
outlined.

\vspace*{3mm}

{\large\bf 2.~~Cross section and kinematics}

\vspace*{3mm}

\begin{figure}
\begin{center}
\includegraphics[width=65mm]{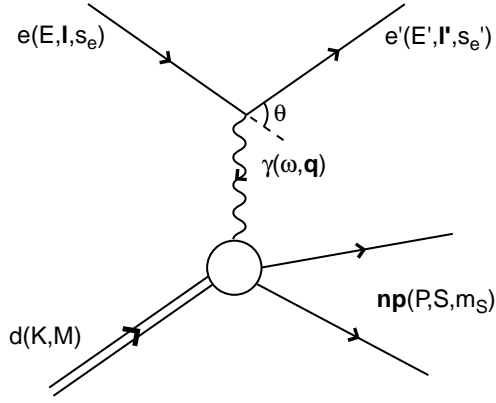}
\caption{\label{OPA}\small One photon approximation.}
\end{center}
\end{figure}

Let us consider the relativistic kinematics of the exclusive
electrodisintegration of the deuteron.  The initial electron
$l=(E,\l)$ collides with the deuteron in rest frame $K=(M_d,\zz)$
($M_d$ is a mass of the deuteron). And there are three particles
in the final state, i.e. electron $l'=(E',\l')$ and pair of proton
and neutron. In one photon approximation (we also neglect the
electron mass) a squared momentum of the virtual photon
$q=(\omega,\q)$ can be expressed via electron scattering angle
$\theta$
\begin{eqnarray}
q^2=-Q^2=(l-l^{\prime})^2=\omega^2-\q^2=-4|\l||\l'|\sin^2\frac{\theta}{2}.
\end{eqnarray}
$np$-pair is described by the invariant mass $s=P^2=(p_p+p_n)^2$
which can be expressed through components of photon 4-impulse:
\begin{eqnarray}
s=M_d^2+2M_d\omega+q^2.
\end{eqnarray}
Lorentz invariant matrix element of the reaction (see
Fig.~\ref{OPA}) can be written as a product of lepton and hadron
currents
\begin{eqnarray}
M_{fi}=-\imath e^2(2\pi)^4\delta^{(4)}(K-P+q) \hskip 60mm
\nonumber\\
\times <l',s_e'|j^{\mu}|l,s_e>
\frac{1}{q^2}<np:(P,Sm_S)|J_\mu|d:(K,M)>,
\end{eqnarray}
where $<l',s_e'|j^{\mu}|l,s_e> = \bar
u(l',s_e')\gamma^{\mu}u(l,s_e)$ is an electromagnetic current
(EM). The initial (final) electron is described by Dirac spinor
$u(l,s_e)$ ($\bar u(l',s_e')$). The hadron current
$<np:(P,Sm_S)|J_\mu|d:(K,M)>$ is a transition matrix element from
the initial deuteron $|d:(K,M)>$ with total momentum $K$,
projection $M$ to the final $np$-pair $|np:(P,Sm_S)>$\ with total
momentum $P$ and spin $S$, projection $m_S$.
Thus the unpolarized cross section of the electrodisintegration of
the deuteron can be easily written as
\begin{eqnarray}
\frac{d^5\sigma}{dE^{\prime}d\Omega^{\prime}d\Omega_\pp}=
\frac{\alpha^2}{8M_d(2\pi)^3}
\frac{|\l'|}{|\l|}\frac{\sqrt s}{q^4}{R}\,
l^{\mu\nu}W_{\mu\nu}, \label{crosssection1}
\end{eqnarray}
where the  factor $R$ connects the final proton angle
in the center of mass system (C.M.S.) (where the $np$-pair is
rest) with the same in the laboratory system (L.S.):
\begin{eqnarray}
R=\frac{\pp^2}{\sqrt{1+\eta}|\pp|-e_\pp\sqrt{\eta}
\cos\theta_{\pp}}.
\end{eqnarray}
Here $\pp$ is a momentum of the final proton in L.S.,
$e_\pp=\sqrt{\pp^2+m^2}$, $\theta_\pp$ is an angle between the
final proton and $Z$-axis, $m$ is a nucleon mass, and
$\eta=\q^2/s$.

Tensor of unpolarized leptons in (\ref{crosssection1}) is expressed as
\begin{eqnarray}
l_{\mu\nu}= \frac12 \sum_{s_e\,s_e^{\prime}}
<l^{\prime},s_e^{\prime}|{{j_{\mu}}^{\dag}}|l,s_e>
<l,s_e|j_{\nu}|l^{\prime},s_e^{\prime}> = 2(l^{\prime}_{\mu}
l_\nu+l^{\prime}_{\nu} l_{\mu} )+g_{\mu\nu}q^2
\end{eqnarray}
and hadron tensor can be written as
\begin{eqnarray}
W_{\mu\nu}= \frac13 \sum_{M S m_s}<d:(K,M)|{J_{\mu}}^{\dag}|np:(P,Sm_S)>
<np:(P,Sm_S)|J_{\nu}|d:(K,M)>.
\end{eqnarray}

In order to average on initial and sum on final states let us
introduce a helicity tensor which can be directly connected with
structure functions (see, for example \cite{arenh3},\cite{gross},\cite{gakh}).
These quantities allow to calculate polarization and asymmetry
observables easily and may be necessary in future (we didn't
calculate them in this work). Keeping in mind the Hermitian
properties of the lepton and the hadron tensors the cross section
can be rewritten as
\begin{eqnarray}
\frac{d^5\sigma}{dE^{\prime}d\Omega^{\prime}d\Omega_\pp}=
\frac{\sigma_{Mott}}{8M_d(2\pi)^3}{\sqrt s}{R}
\hskip 70mm
\nonumber\\
\times\left[ l^0_{00}W_{00}+l^0_{++}(W_{++}+W_{--})+ l^0_{+-}2 \re
W_{+-}-l^0_{0+}2 \re (W_{0+}-W_{0-})\right], \label{crosssection2}
\end{eqnarray}
where $\sigma_{Mott} = (\alpha \cos\frac{\theta}{2}/
2E\sin^2\frac{\theta}{2})^2$ is Mott cross section for point-like particles
and
\begin{eqnarray}
l_{00}^0=\frac{Q^2}{\q^2},\quad l_{0+}^0=\frac{Q}{|\q|\sqrt
2}\sqrt{\frac{Q^2}{\q^2}+\textmd{tg}^2 \frac{\theta}{2}},\quad
l_{++}^0=\frac{1}{2}\textmd{tg}^2\frac{\theta}{2}+\frac{Q^2}{4\q^2},\quad
l_{+-}^0=-\frac{Q^2}{2\q^2}.
\end{eqnarray}
So the calculation of the cross section (\ref{crosssection2})
comes to the calculation of the hadron tensor $W_{\mu\nu}$ which
describes the NN interaction and is a subject of our
investigation.

\vspace*{3mm}
{\large\bf 3.~~Bethe-Salpeter Amplitude of the Deuteron}
\vspace*{3mm}

In the Bethe-Salpeter approach (BSA) the deuteron is a bound system
which can be described by the amplitude $\Phi_{M}(k;K)$ of the equation
\begin{eqnarray}
\Phi_{M\, \alpha\beta}(k;K) = \imath
S^{(1)}_{\alpha\eta}\left(\frac{K}{2}+k\right)
S^{(2)}_{\beta\rho}\left(\frac{K}{2}-k\right) \int
\frac{d^4k^{\prime}}{(2\pi)^4}
V_{\eta\rho,\epsilon\lambda}(k,k^\prime;K) \Phi_{M\,
\epsilon\lambda}(k^\prime;K), \label{bsamplitude}
\end{eqnarray}
here $S^{(\ell)}(K/2 - (-1)^{\ell} k)$ is a propagator of the
$\ell$-th nucleon, $V(k,k^\prime;K)$ is a kernel of a NN
interaction (Greek letters means spinor indexes). The amplitude of
the deuteron in rest frame can be expanded through two-nucleon
relativistic states $|aM\rangle\equiv|\pi,\,{}^{2S+1}L_J^{\rho} M
\rangle$:
\begin{eqnarray}
\Phi_{M}(k;K)=\sum_{a} \phi_a(k_0,|\k|) {\cal
Y}_{aM}(\k),\label{partial}
\end{eqnarray}
where $S$ denotes a total spin of the system, $L$ is an orbital
angular momentum, $J$ is a total angular momentum with a
projection $M$. Quantum number $\rho$ counts positive- and
negative-energy states, $\pi$ marks the parity of the state;
$\phi_a(k_0,|\k|)$ is the radial part of the BS amplitude. The
spin-angular part ${\cal Y}_{aM}(\k)$) is
\begin{eqnarray}
{\cal Y}_{aM}(\k) U_C =
\hskip 96mm\nonumber\\
\imath^{L} \sum_{m_Lm_Sm_1m_2\rho_1\rho_2} C_{\frac12 \rho_1
\frac12 \rho_2}^{S_{\rho} {\rho}} C_{L m_L S m_S}^{JM} C_{\frac12
m_1 \frac12 m_2}^{Sm_S} Y_{L{m_L}}(\k) {u^{\rho_1}_{m_1}}(\pp)
{{u^{\rho_2}_{m_2}}}^{T}(-\k).
\end{eqnarray}

Using partial wave decomposition (\ref{partial}) we can write the
decomposed Bethe-Salpeter equation for the radial part of the BSA
\begin{eqnarray}
\phi_{a}(k_0,|\k|) = S_{a}(k_0,|\k|;s) \int d k_0^{\prime} \int
{\k^{\prime}}^2 d|\k^{\prime}| \sum\limits_{b} V_{a
b}(k_0,|\k|,k_0^{\prime},|\k^{\prime}|;s)
\phi_{b}(k_0^{\prime},|\k^{\prime}|).\label{BSsepar}
\end{eqnarray}
Then taking into account separable {\em anzats}
for the rank $N$ kernel of interaction
\begin{eqnarray}
V_{a b}(k_0,|\k|,k_0^{\prime},|\k^{\prime}|;s) =
\sum\limits_{i,j=1}^{N} \lambda_{ij}\, g_i^{a}(k_0,|\k|)\,
g_j^{b}(k_0^{\prime},|\k^{\prime}|),\qquad
\lambda_{ij}=\lambda_{ji},
\end{eqnarray}
we can find  a solution of the BS equation (\ref{BSsepar})
in the following form
\begin{eqnarray}
\phi_{a}(k_0,|\k|) = \sum\limits_{i,j=1}^{N}
S_{a}(k_0,|\k|;s)\,\lambda_{ij}\,g_{i}^{a}(k_0,|\k|)\,c_j(s).
\end{eqnarray}
Here $\lambda_{ij}$ are fitting parameters, $g_i^{a}(k_0,|\k|)$ are trial functions.
The coefficients $c_j(s)$ satisfy the following system of homogeneous equations
\begin{eqnarray}
c_i(s)\,-\,\sum\limits_{k,j=1}^{N}\,h_{ik}(s)\,\lambda_{kj}\,c_j(s)\,=\,0,
\end{eqnarray}
where $h_{ik}(s)$ are defined by the integral
\begin{eqnarray}
h_{ik}(s) = \frac{\imath}{2\pi^2}\sum\limits_{a} \int\,d k_0\,\int
\k^2\, d|\k|\,S_{a}(k_0,|\k|;s)\,g_i^{a}(k_0,|\k|)\,
g_k^{a}(k_0,|\k|).
\end{eqnarray}
Using the covariant separable Graz II rank III kernel of
interaction \cite{burov} we can find $\lambda_{ij}$ (see Table
\ref{tab:graz}) from an analysis of the experimental data for the
deuteron characteristics
(phase shifts, binding energy, length of scattering etc.).

\begin{table}[th]
\caption{\label{tab:graz}\small Parameters of the covariant
separable Graz II kernel} {\[
\begin{tabular}{|lrll|lrll|}
\hline $\gamma_1$     & 28.69550 & &GeV$^{-2}$ &
$\lambda_{11}$ & 2.718930 &$\times$ 10$^{-4}$ & GeV$^{6}$ \\
$\gamma_2$     & 64.9803  &                   & GeV$^{-2}$ &
$\lambda_{12}$ & -7.16735 &$\times$ 10$^{-2}$ & GeV$^{4}$ \\
$\beta_{11}$   & 2.31384  &$\times$ 10$^{-1}$ & GeV        &
$\lambda_{13}$ & -1.51744 &$\times$ $10^{-3}$ & GeV$^{6}$ \\
$\beta_{12}$   & 5.21705  &$\times$ $10^{-1}$ & GeV        &
$\lambda_{22}$ & 16.52393 &                   & GeV$^{2}$ \\
$\beta_{21}$   & 7.94907  &$\times$ $10^{-1}$ & GeV        &
$\lambda_{23}$ & 0.28606  &                   & GeV$^{4}$ \\
$\beta_{22}$   & 1.57512  &$\times$ $10^{-1}$ & GeV        &
$\lambda_{33}$ & 3.48589  &$\times$ $10^{-3}$ & GeV$^{6}$ \\
\hline
\end{tabular}\label{lambda}
\]}
\end{table}

It is necessary to remark that here we took into account the
positive-energy states ($^3S_1^+$, $^3D_1^+$) only
\begin{eqnarray}
&&g_{1}^{^3S_1^+}(k_0,|\k|)=\frac{1-\gamma_{1}(k_0^2-\k^2)}
{(k_0^2-\k^{2}-\beta_{11}^{2})^{2}}\\
&&g_{2}^{^3S_1^+}(k_0,|\k|)=-\frac{(k_0^2-\k^2)}
{(k_0^2-\k^2-\beta_{12}^2)^{2}}\\
&&g_{3}^{^3D_1^+}(k_0,|\k|)=\frac{(k_0^2-\k^{2})
(1-\gamma_{2}(k_0^2-\k^{2}))}
{(k_0^2-\k^{2}-\beta_{21}^{2})(k_0^2-\k^{2}-\beta_{22}^{2})^{2}}\\
&&g_{1}^{^3D_1^+}(k_0,|\k|)=g_{2}^{^3D_1^+}(k_0,|\k|)=
g_{3}^{^3S_1^+}(k_0,|\k|)\equiv 0.
\end{eqnarray}

In our calculation it is more convenient to use the BS vertex
$\Gamma_M$ which is related with the BS amplitude by simple
expression. For full functions
\begin{eqnarray}
\Phi_M(k;K)= S^{(1)}\left(\frac{K}{2}+k\right)
S^{(2)}\left(\frac{K}{2}-k\right)\Gamma_M(k;K),
\end{eqnarray}
and for their radial parts
\begin{eqnarray}
\phi_a(k_0,|\k|)=\sum_{b} S_{ab}(k_0,|\k|;s) g_{b} (k_0,|\k|),
\end{eqnarray}
here we omitted spinor indexes for simplicity. $S_{ab}$ is a
nucleon propagator which is diagonal for positive-energy partial
parts
\begin{eqnarray}
S_{++}(k_0,|\k|;s)= 1/\left[\left(\sqrt{s}/{2}+k_0-e_\k\right)
\left(\sqrt{s}/{2}-k_0-e_\k\right)\right].
\end{eqnarray}

\vspace*{3mm}

{\large\bf 4.~~Hadron Electromagnetic Current}

\vspace*{3mm}

Let us write the matrix element of the hadron electromagnetic
current with the BS amplitude using Mandelstam technique
\cite{mandelstam}
\begin{eqnarray}
<np:(P,Sm_S)|J_\mu|d:(K,M)>= \hskip 70mm
\nonumber\\
\imath\int \frac{d^4p}{(2\pi)^4}\,\frac{d^4k}{(2\pi)^4}\,
\bar\chi_{Sm_s}(p;p^{*},P)\,\Lambda_{\mu}(p,k;P,K)\,\Phi_M(k;K).
\end{eqnarray}
Here $p^*$ is the relative 4-momentum of the $np$-pair (nucleons
are on-mass-shell), $Pp^*=0$, and ${\pp^*}^2=s/4-m^2$.

\begin{figure}
\begin{center}
\includegraphics[width=65mm]{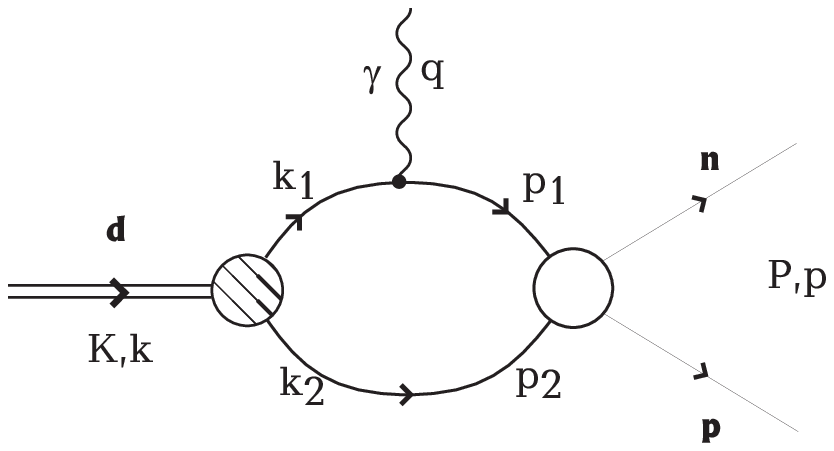}~~~~~~~~~~~~~~~~~
\includegraphics[width=65mm]{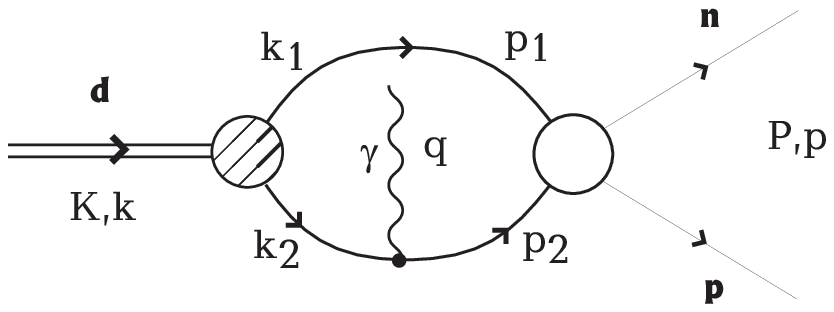}
\caption{\label{kinematics}\small Relativistic impulse approximation.}
\end{center}
\end{figure}

We consider the process of the electrodisintegration of the
deuteron in RIA (see Fig. \ref{kinematics}).
In our further calculation only one-body currents are taken into
account
\begin{eqnarray}
\Lambda^{[1]}_{\mu}(p,k;P,K) = \imath(2\pi)^4 \left\{
\delta^{(4)}(p-k-\frac{q}{2})
\Gamma^{(1)}_\mu\left(\frac{P}{2}+p,\frac{K}{2}+k\right)
{S^{(2)}\left(\frac{P}{2}-p\right)}^{-1} \right. \hskip 5mm
\nonumber\\
\left. +\delta^{(4)}(p-k+\frac{q}{2})
\Gamma^{(2)}_\mu\left(\frac{P}{2}-p,\frac{K}{2}-k\right)
{S^{(1)}\left(\frac{P}{2}+p\right)}^{-1} \right\}
\end{eqnarray}
(here the total and relative momenta are introduced: $P=p_1+p_2,
K=k_1+k_2, k=\frac{1}{2}(k_1-k_2), p=\frac{1}{2}(p_1-p_2)$).

In this case the matrix element of the hadron current has the
following form
\begin{eqnarray}
<np:(P,Sm_S)|J_\mu|d:(K,M)>= \imath \sum_{\ell=1,2} \int
\frac{d^4p}{(2\pi)^4}\ \bar\chi_{Sm_S}(p;P)
\nonumber\\
\Gamma_\mu^{(\ell)}(q)\
S^{(\ell)}\left(\frac{P}{2}-(-1)^{\ell}p-q\right)\
\Gamma_{M}\left(p+(-1)^{\ell}\frac{q}{2};K\right).
\label{emcurrent1}
\end{eqnarray}
Note that $\gamma NN$- vertex was taken on-mass-shell
\begin{eqnarray}
\Gamma_\mu^{(\ell)}(p',p)\longrightarrow\Gamma^{(\ell)}_\mu(q)=\gamma_\mu
F^{(\ell)}_1(q^2) -\frac{1}{4m}\left(\gamma_\mu\hat q-\hat
q\gamma_\mu\right)F^{(\ell)}_2(q^2).
\label{gammaNN}
\end{eqnarray}
Here $F^{(\ell)}_1$ ($F^{(\ell)}_2$) - Dirac (Pauli) form factor
of the nucleon which obeys the next normalization conditions
\begin{eqnarray}
F^{(1)}_1(0) = 1,\qquad F^{(1)}_2(0) = \varkappa_p,
\nonumber\\
F^{(2)}_1(0) = 0,\qquad F^{(2)}_2(0) = \varkappa_n,
\end{eqnarray}
$\varkappa_p$ ($\varkappa_n$) is the anomalous proton (neutron)
magnetic moment.

It should be mentioned that we neglect a final state interaction
(FSI) (it is a subject of future calculations).
\begin{eqnarray}
&&\bar\chi^{(0)}_{Sm_s}(p;p^{*},P) =
(2\pi)^4\,\delta^{(4)}(p-p^*)\,\bar\chi^{(0)}_{Sm_S}(p^*,P)
\nonumber\\
&&= (2\pi)^4\,\delta^{(4)}(p-p^*)\,\sum_{m_1m_2}
C_{\frac{1}{2}m_1\frac{1}{2}m_2}^{Sm_S}\, {\bar
u}_{m_1}\left(\frac{P}{2}+p\right)\, {\bar
u}_{m_2}\left(\frac{P}{2}-p\right). \label{pwa}
\end{eqnarray}

\begin{figure}[t]
\begin{center}
\includegraphics[width=120mm]{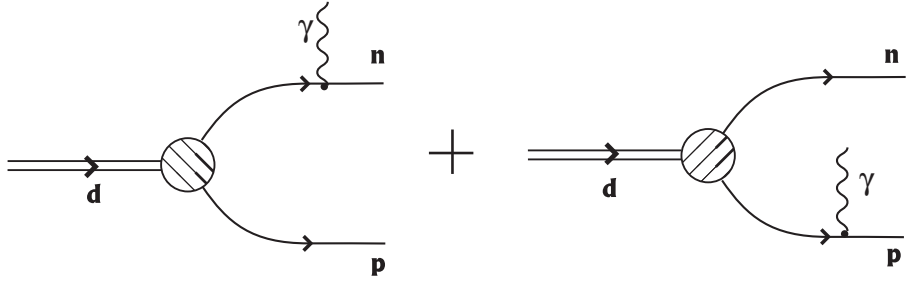}
\caption{\label{pwaria}\small Plane wave approximation.}
\end{center}
\end{figure}

Using (\ref{pwa}) for
the final $np$-pair wave function and integrating
(\ref{emcurrent1}) over $p$ we obtain our basic RIA expression for
the hadron current
\begin{eqnarray}
<np:(P,Sm_S)|J_\mu|d:(K,M)>= \imath \sum_{\ell=1,2}
\bar\chi^{(0)}_{Sm_S}(p^*,P) \Gamma_\mu^{(\ell)}(q)
\nonumber\\
\times S^{(\ell)}\left(\frac{K}{2}-p^*-(-1)^\ell\frac{q}{2}\right)
\Gamma_{M} \left(p^*+(-1)^\ell\frac{q}{2};K \right).
\label{emcurrent2}
\end{eqnarray}
It has very simple form. And to get it we should just perform the
analytical calculation of the trace. For this purpose we use the REDUCE
system.

\vspace*{3mm}

{\large\bf 5.~~Factorization of the cross section}

\vspace*{3mm}

Let us consider the electrodisintegration of the deuteron
supposing that an initial lepton collides only with the proton in
the deuteron and the neutron is a spectator. Then the cross
section is factorized on two parts, one is connected with the
contribution of the neutron as a spectator and another with a
proton contribution, the latter does not have interference terms
between the $S$- and $D$-states.

\vspace*{3mm}

{\bf 5.1~~Nonrelativistic case}

\vspace*{3mm}

The amplitude of the process can formally be presented
as a production
\begin{eqnarray}
{\cal M}=\chi^+_{m_1}\chi^+_{m_2}\hat O\Psi_M,
\end{eqnarray}
where spinors $\chi^+_{m_1},~\chi^+_{m_2}$ describe the outgoing
$np$-pair, $\hat O$ corresponds to the interaction vertex,
$\Psi_M$ is a wave function of the deuteron. Let us note that the
vertex $\hat O$ stands in general for any one-particle
interaction, but in this paper describes $\gamma N N$-vertex.

Inserting into this expression a complete set of pair states we
can get
\begin{eqnarray}
{\cal M}_{\mu}=\sum_{m_1'}(\chi^+_{m_1}\hat
O_{\mu}\chi_{m_1'})\chi^+_{m_1'}\chi^+_{m_2}\Psi_M,
\end{eqnarray}
after evident transformations the hadron tensor can be written as
\begin{eqnarray}
W_{\mu\nu}= \frac13 \sum_{m_1 m_2 M} {\cal M}_{\mu}{\cal M}_{\nu}
&=& \frac{1}{3}\sum_{m_1m_2M}\left|\sum_{m_1'}\left[\chi^+_{m_1}
\hat O\chi_{m_1'}\right]\left[\chi^+_{m_1'} \chi^+_{m_2}
\Psi_M\right]\right|^2\nonumber\\
&=&\frac{1}{3}\sum_{m_1m_2M} \sum_{m_1'm_1''}\left[
\chi_{m_1}^+\hat O\chi_{m_1'}\right]^2\left[\chi^+_{m_1'}
\chi_{m_2}^+\Psi_M\Psi^+_M\chi_{m_2}\chi_{m_1''}\right].
\nonumber
\end{eqnarray}
Introducing the partial-wave decomposition for the deuteron:
$$\Psi_M =\sum_{lm s_1 s_2 s} C_{lm 1 s}^{1M}
C_{\frac{1}{2}s_1\frac{1}{2}s_2}^{1 s}Y_{lm}
\chi_{s_1}\chi_{s_2}u_l$$ and transforming the second term with
the help of orthogonalization properties of the spinor $\chi$ and
some relations for Clebsh-Gordan coefficients we can finally
obtain the factorized expression
\begin{eqnarray}
W_{\mu\nu} = \frac{1}{8\pi}|A_{\mu}A^*_{\nu}| \sum_l|u_l|^2,
\end{eqnarray}
where $A_{\mu}=\chi_{m_1}^+\hat O_{\mu}\chi_{m_1'}$ is a one-body
interaction part and $l$ counts partial states of the deuteron.

Thus it is seen that the cross section is proportional to the sum
of squared radial parts of the deuteron wave function.

\vspace*{3mm}

{\bf 5.2~~Relativistic case}

\vspace*{3mm}

In the relativistic case the matrix element of the deuteron
electrodisintegration can be written schematically in the
following form
\begin{eqnarray}
{\cal M}=\Psi_{pair}\otimes\hat O\otimes S\otimes\Gamma_M,
\end{eqnarray}
where $\Psi_{pair}$ is the wave function of the $np$-pair,
$\hat O$ is a vertex of interaction,
$S$ is a propagator of the nucleon,
$\Gamma_M$ is the vertex function of the deuteron.

Introducing the partial-wave decomposition of the deuteron vertex
function in the L.S., considering the only proton interacting with
a virtual photon and supposing the PWA for the final $np$-pair we
can present the comprehensive expression in the following form
\begin{eqnarray}
{\cal M}_{\mu}&=&\sum_{s_1s_2}C_{\frac{1}{2}s_1\frac{1}{2}
s_2}^{SM_S} \bar u^{(1)}(s_1,\pp_1)\bar u^{(2)}(s_2,\pp_2)
\Gamma^{(1)}_\mu(q)
S_{1\,+}(\k_1) \sum_{m_1'}u^{(1)} (m_1',\k_1)
\bar u^{(1)}(m_1',\k_1)
\nonumber\\
&&\times\sum_{m_1m_2lmsm_s}C_{lmsm_s}^{1M}
C_{\frac{1}{2}m_1\frac{1}{2}m_2}^{sm_s}u^{(1)}
(m_1,-\k_1)u^{(2)}(m_2,-\k_2)Y_{lm}(\hat\k) g_{l}(k_0,|\k|),
\nonumber
\end{eqnarray}
where $S_{1\,+}(\k_1) = 1/(k_{10}-e_{\k_1})$ and $\Gamma^{(1)}_\mu(q)$
vertex is described by Eq.~(\ref{gammaNN}). As it was assumed
above only $^3S_1^+,~^3D_1^+$-states were taken into account.
Using orthogonalization properties of the bispinors and some
relations for Clebsh-Gordan coefficients we can write
\begin{eqnarray}
{\cal M}_{\mu} = \sum_{s_1s_2m_1lmsm_s}
C_{\frac{1}{2}s_1\frac{1}{2}s_2}^{SM_S}C_{lmsm_s}^{1M}
C_{\frac{1}{2}m_1\frac{1}{2}s_2}^{sm_s}Y_{lm}(\hat\k)
g_{l}(k_0,|\k|) S_{1\,+}(\k_1)
A^{(1)}_{\mu}(s_1,\pp_1;m_1,\k_1),
\nonumber
\end{eqnarray}
with
\begin{eqnarray}
A^{(1)}_{\mu}(s_1,\pp_1;m_1,\k_1)=\bar u^{(1)}(s_1,\pp_1)
\Gamma^{(1)}_\mu(q) u^{(1)}(m_1,\k_1),
\end{eqnarray}
is a one-body photon-proton interaction part. Now we can
derive the hadron tensor
\begin{eqnarray}
W_{\mu\nu}=\frac{1}{3}\sum_{MSM_S}{\cal M}_{\mu}{\cal M}_{\nu}.
\nonumber
\end{eqnarray}
Using once more properties of the Clebsh-Gordan coefficients and
Dirac spinors we obtain the expression
\begin{eqnarray}
W_{\mu\nu}= C_d\ Sp \left\{(\hat p_1+m) \Gamma^{(1)}_\mu(q) (\hat
k_1+m) {\bar \Gamma^{(1)}}_\mu(q)\right\} \label{Wspur}
\end{eqnarray}
which involves the simply calculated trace and a function
$$C_d = \frac{1}{8\pi}\frac{1}{4e_{\k_1}e_{\pp_1}}S_{1\,+}^2(\k_1)
\sum_{l=0,2} |g_{l}(k_0,|\k|)|^2$$ containing the structure of the
deuteron. Performing the trace calculation we finally obtain the
expression for the hadron tensor
\begin{eqnarray}
W_{\mu\nu}= C_d\ \left( W^a_{\mu\nu} F_1^2(q^2) + W^b_{\mu\nu}
F_1(q^2)F_2(q^2) + W^c_{\mu\nu} F_2^2(q^2) \right)
\end{eqnarray}
with
\begin{eqnarray}
&&W^a_{\mu\nu} = 4\left[p_{1\,\mu} k_{1\,\nu}+k_{1\,\mu} p_{1\,\nu}
+(m^2-(p_1\cdot k_1))g_{\mu\nu}\right]
\label{Wparts}\\
&&W^b_{\mu\nu} =  2\left[k_{1\,\mu} q_{\nu}-q_{\mu} k_{1\,\nu}
-p_{1\,\mu} q_{\nu}+q_{\mu} p_{1\,\nu} \right]
\nonumber\\
&&W^c_{\mu\nu} =
\left[\left(-q^2 m^2-q^2(p_1\cdot k_1)+2(p_1\cdot q)(k_1\cdot q)\right)
g_{\mu\nu} + \left(m^2+(p_1\cdot k_1)\right)q_\mu q_\nu\nonumber\right.\\
&&-\left.\left((k_1\cdot q)(p_{1\,\mu} q_{\nu}+q_{\mu}p_{1\,\nu})
+(p_1\cdot q)(k_{1\,\mu} q_{\nu}+q_{\mu} k_{1\,\nu})
-q^2(p_{1\,\mu}
k_{1\,\nu}+k_{1\,\mu}p_{1\,\nu})\right)\right]/m^2.
\nonumber\end{eqnarray} Let us note here in the expressions
Eqs.~(\ref{Wspur},\ref{Wparts}) the four-vector $k_1$ has the
on-mass-shell form $k_1 = (e_{\k_1},\k_1)$ in differ with $k_1 =
(k_{10},\k_1)$ in the Fig.~\ref{kinematics}.

Thus we see that the factorization of the electrodisintegration
cross section exists both in nonrelativistic and relativistic
cases. The necessary conditions for this are the plane-wave
approximation for the final $np$-pair, the neutron in the deuteron
is supposed to be a spectator (the one-body type of the
interaction in the vertex ${\hat O}$) and only positive-energy
states for the deuteron are taking into account. As for the second
condition the type of one-body interaction does not play any role
but only spin-one-half particle is scattered. The third condition
means the $P$ waves in the deuteron (namely $^3P_1^{+-}$ and
$^1P_1^{+-}$) destroy the factorization.

\vspace*{3mm}

{\large\bf 6.~~Results and Discussion}

\vspace*{3mm}

We present here the results of the calculation of
the deuteron electrodisintegration cross section in the
relativistic plane wave impulse approximation with the separable
Graz II rank III kernel of interaction. In our calculations we
follow to conditions of real experiments and we distinguish eight
sets of experimental data. Let us mark these sets as $Saclay_I$,
$Saclay_{II}$ (see \cite{bernheim}, Table 3);  $Saclay_{III}$
(\cite{turck}, Table 1); $Bonn_{I}$, $Bonn_{II}$ (see
\cite{breuker}, Table 3); $Bonn_{III}$, $Bonn_{IV}$, $Bonn_{V}$
(see \cite{boden}, Tables 5,3,4 respectively).

\begin{figure}
\begin{center}
\includegraphics[width=110mm]{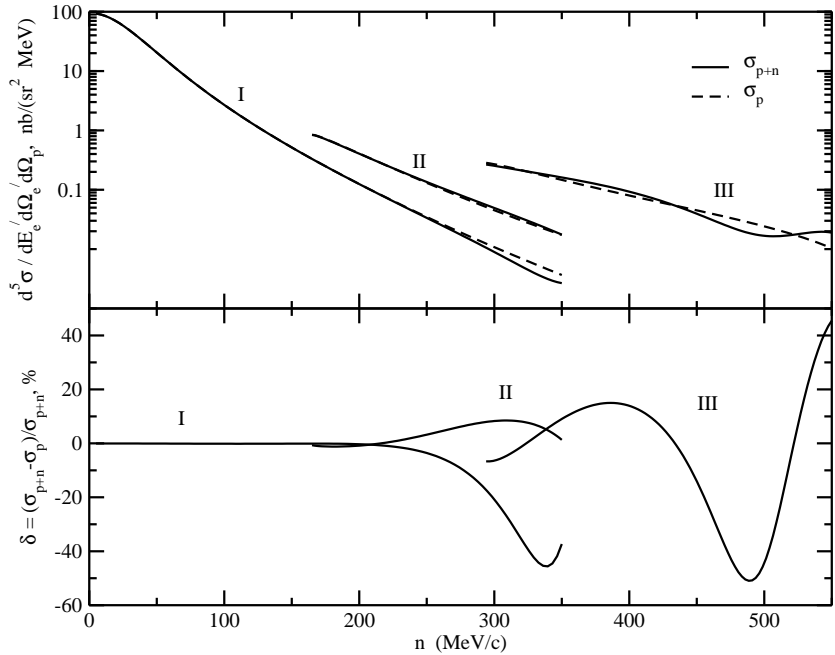}
\caption{{\small The electrodisintegration cross section versus
the neutron momentum for three kinematics of the experiments at
Saclay. Solid and dashed lines correspond to the calculations with
and without neutron contribution (upper plot). Bottom plot shows
the relative neutron contribution in the corresponding
experimental regions. The experimental data regions were taken
from \protect\cite{bernheim}($Saclay_{I,II}$) and
\cite{turck}($Saclay_{III}$).}} \label{dkin_123c_bp_n}
\end{center}
\end{figure}

\begin{figure}
\begin{center}
\includegraphics[width=90mm]{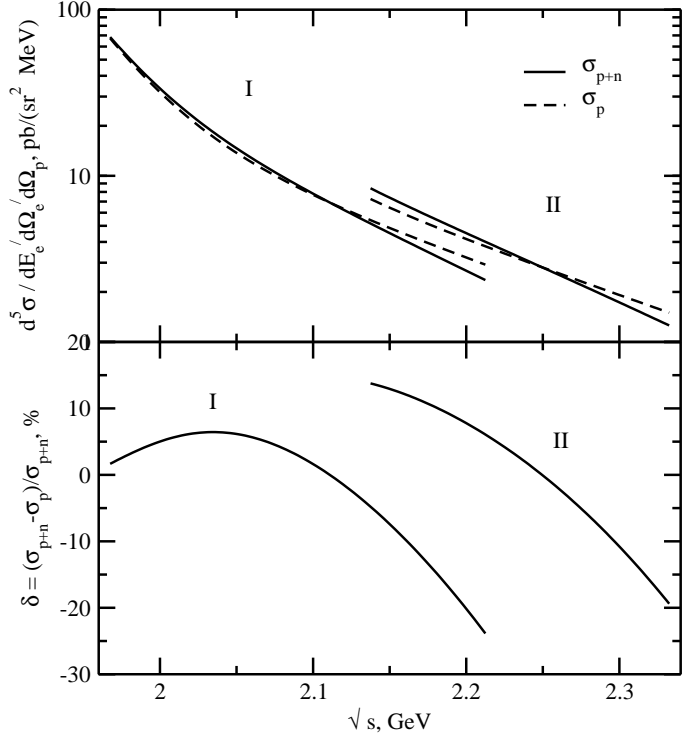}
\caption{{\small The same as in previous figure but versus pair
invariant mass $\sqrt{s}$ for the kinematical conditions were
taken from \cite{breuker} ($Bonn_{I,II}$).}}
\label{dkin_s12c_bp_n}
\end{center}
\end{figure}

\begin{figure}
\begin{center}
\includegraphics[width=90mm]{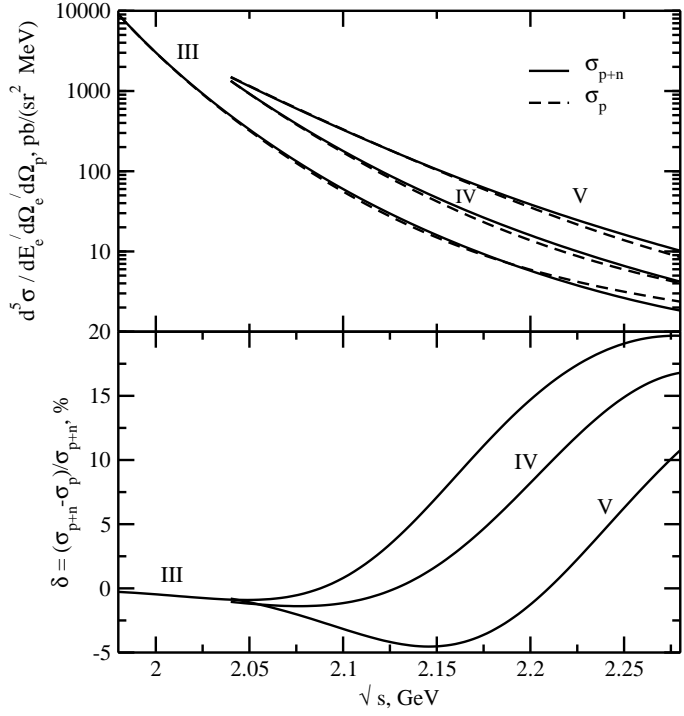}
\caption{{\small The same as in previous figure. The kinematical
conditions were taken from \cite{boden}($Bonn_{III,IV,V}$).}}
\label{dkin_s3_123c_bp_n}
\end{center}
\end{figure}

\begin{figure}
\begin{center}
\includegraphics[width=110mm]{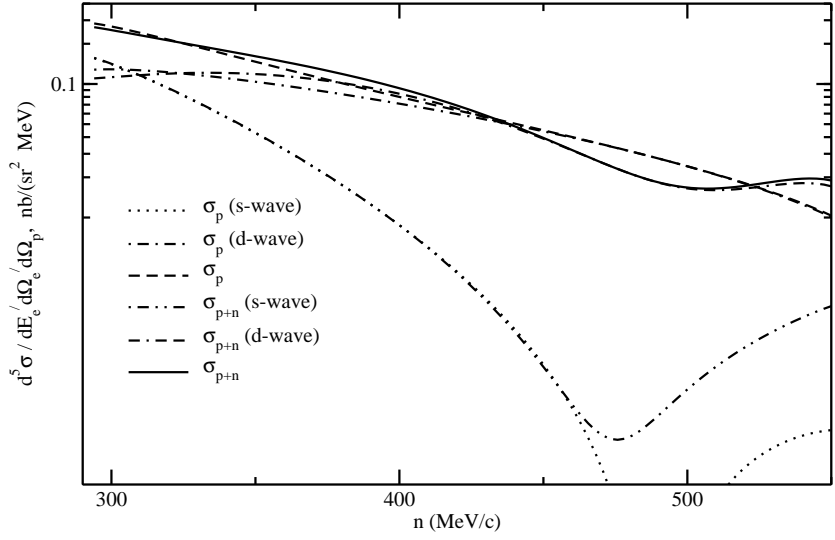}
\caption{{\small The contributions of the spectator neutron versus
the outgoing neutron momentum to the electrodisintegration cross
section of the deuteron partial $S$-, $D$-states are shown for the
experimental sets from \cite{turck}($Saclay_{III}$).}}
\label{dkin_3c_bp}
\end{center}
\end{figure}

\begin{figure}
\begin{center}
\includegraphics[width=110mm]{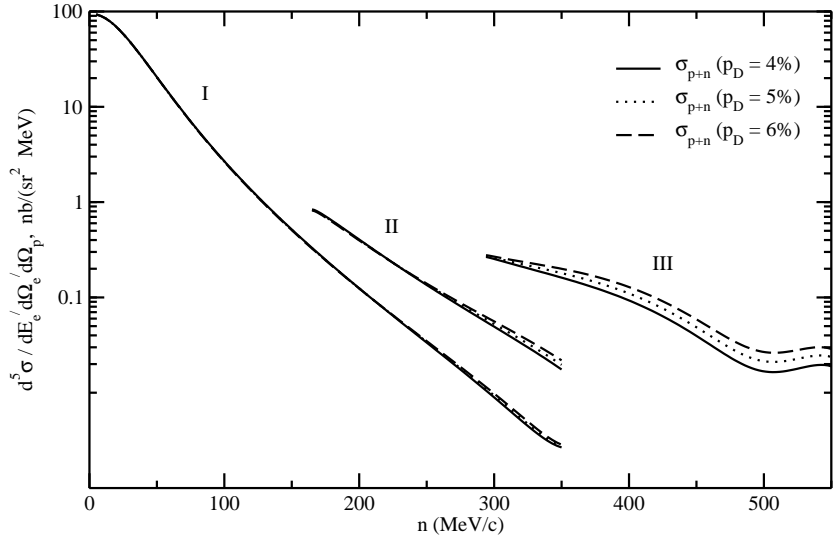}
\caption{{\small The contributions of the deuteron partial D-state
to the electrodisintegration cross section versus neutron momenta
are shown for three sets of the experiments
\cite{bernheim}($Saclay_{I,II}$), \cite{turck} ($Saclay_{III}$).}}
\label{dkin_123c_pd}
\end{center}
\end{figure}

\begin{figure}
\begin{center}
\includegraphics[width=110mm]{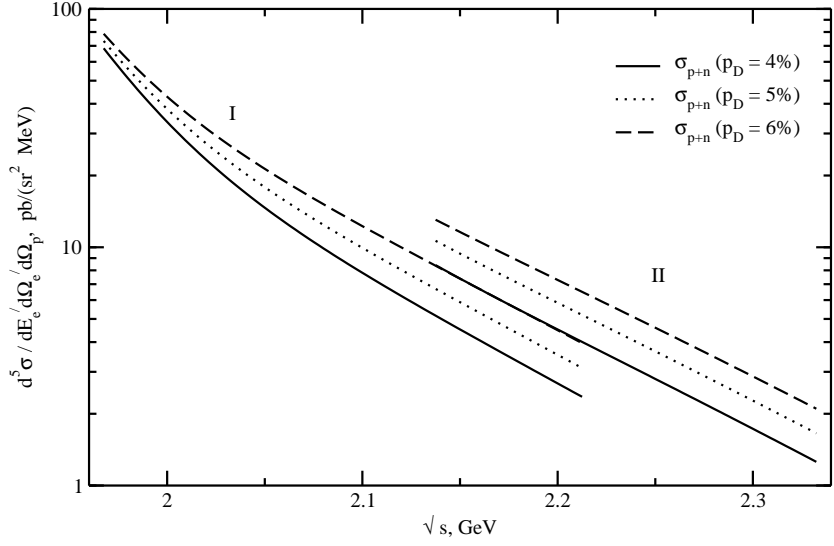}
\caption{{\small The contributions of the deuteron partial D-state
to the electrodisintegration cross section versus pair invariant
mass $\sqrt{s}$ are shown for the conditions of the experiments
\cite{breuker} ($Bonn_{I,II}$).}} \label{dkin_s12c_pd}
\end{center}
\end{figure}

\begin{figure}
\begin{center}
\includegraphics[width=110mm]{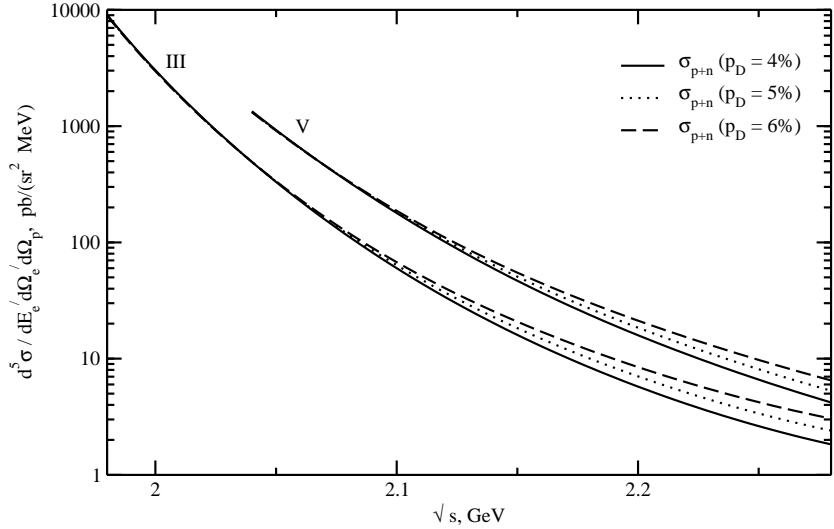}
\caption{{\small The same as in the previous figure but for
\cite{boden} conditions ($Bonn_{III,V}$). We omitted curves for
the $Bonn_{IV}$ case because they are very close to $Bonn_{V}$.}}
\label{dkin_s3_123c_pd}
\end{center}
\end{figure}

\begin{figure}
\begin{center}
\includegraphics[width=110mm]{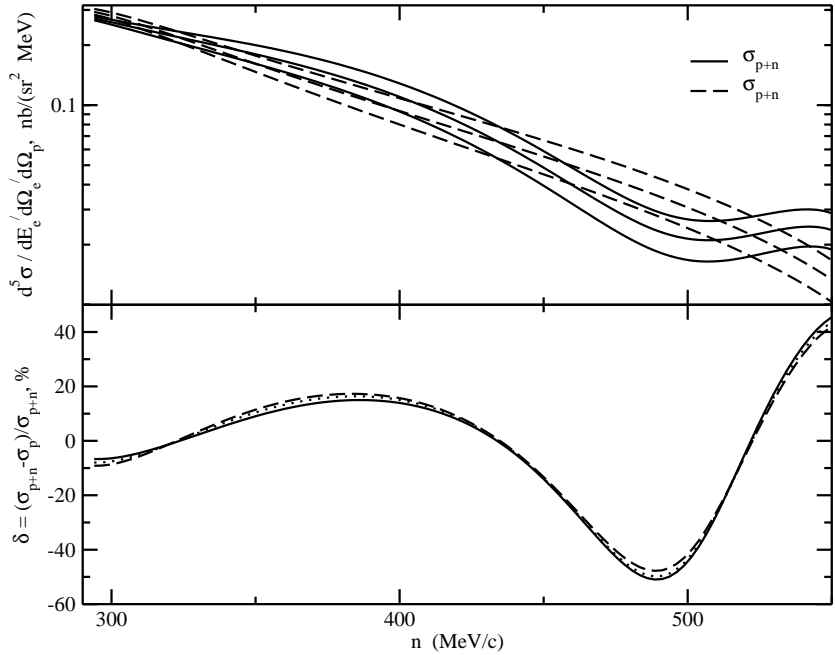}
\caption{{\small The contribution of the spectator neutron versus
neutron momenta to the electrodisintegration cross section for
different deuteron $D$-states for conditions of the experiment
$Saclay_{III}$ \cite{turck}. In the first picture solid (dashed)
line stands for $p+n$- ($p$-) contribution with different
$D$-states in the deuteron: $p_D$ = 4\% for lower line and $p_D$ =
6\% for upper line.}} \label{dkin_3c_bpd}
\end{center}
\end{figure}

First of all we illustrated the influence of the spectator neutron
on the cross section (see Figs. \ref{dkin_123c_bp_n},
\ref{dkin_s12c_bp_n}, \ref{dkin_s3_123c_bp_n}). It is seen that it
increases with the increasing of the neutron momentum and reaches
50\% in the $Saclay_{III}$ kinematic range. One can see that the
cross section of the deuteron electrodisintegration
versus $\sqrt{s}$ changes not so strong, nevertheless  the
contribution of the spectator neutron is not negligible. Let us
note that this contribution changes sign in the $Saclay_{III}$
kinematic region (see Fig. \ref{dkin_123c_bp_n}). In order to
understand the origin of this behavior we present on the Fig.
\ref{dkin_3c_bp} partial contributions of the S- and D-states for
this  kinematical region versus neutron momenta.
We found that the D-state plays an important role and then it is
naturally to ask what happens if we change the magnitude of the
D-state. On the Figs. \ref{dkin_123c_pd}, \ref{dkin_s12c_pd},
\ref{dkin_s3_123c_pd}
we can see that for different magnitudes of the D-states the cross
section changes distinctly (especially for the kinematic region
\cite{turck}), but the ratio $\delta =
(\sigma_{p+n}-\sigma_p)/\sigma_{p+n}$ is not changed at all (see
Fig. \ref{dkin_3c_bpd}). It means that the difference is mainly
connected with the spectator neutron contribution. Thus we can
make a conclusion that the experimental data within the kinematics
from $Saclay_{III}$  \cite{turck} can supply the good test 
for various models of NN interactions in the deuteron.

{\large\bf 7.~~Summary}

\vspace*{3mm}

In the presented paper we have considered the
electrodisintegration of the deuteron in the Bethe-Salpeter
approach. It is realized for the two-nucleon system by using the
multipole expansion with the spinor structure of two nucleons. The
separable ansatz for the interaction kernel
has provided a manageable system of linear homogeneous equations
for deriving the BS amplitude.

We have switched then to the case with the use of the covariant
revision of the Graz II separable potential with the summation of
several separable functions.

The reaction of the deuteron electrodisintegration served as a
testing ground for the method under investigation and helped to
outline both strong and weak points of the approach. The analysis
has proved the technique to be very promising, even if we find an
evident discrepancies with experimental data at this stage of development.
Several items can be suggested for the program of further
theoretical study. First of all it is necessary to take into
account the final state interaction for the $np$-pair. Then we
need to take into account the negative-energy states for the BS
amplitude and calculate the contribution of the $P$ waves in the
electrodisintegration. After that we can calculate different
asymmetries of the $(ed\to enp)$ process which can give new
qualitative information about the structure of the deuteron.

\newpage

{\large\bf Acknowledgments}

\vspace*{3mm} We wish to thank our collaborators K.Yu.~Kazakov,
A.V.~Shebeko, S.Eh.~Shirmovsky, D.V. Shulga for their contribution
to the presented work. We would like to thank Professor H.Toki and
Professor D.~Blaschke for their interest to this work and fruitful
discussions.

\vspace*{3mm} The work is supported in part by the Russian
Foundation for Basic Research, grant No.05-02-17698a.

\end{document}